\documentclass[aps,prd,amsmath,floats,floatfix,twocolumn,superscriptaddress,
  nofootinbib,showpacs]{revtex4}

\usepackage{amssymb}
\usepackage{siunitx}

\usepackage{epsfig}

\newcommand{\Cornell}{\affiliation{Center for Radiophysics and Space
    Research, Cornell University, Ithaca, New York 14853, USA}}
\newcommand{\CITA}{\affiliation{Canadian Institute for Theoretical
    Astrophysics, 60 St.~George Street, University of Toronto,
    Toronto, ON M5S 3H8, Canada}}
\newcommand{\LBL}{\affiliation{Lawrence Berkeley National Laboratory,
1 Cyclotron Rd, Berkeley, CA 94720, USA; Einstein Fellow}}

\newcommand{\SpEC}{\textsc{SpEC}}
\newcommand{\Msun}{\text{\ensuremath{M_\odot}}}

\renewcommand{\Re}{\operatorname{Re}}
\renewcommand{\Im}{\operatorname{Im}}

\newcommand{\fig}[1]{Fig.~\ref{fig:#1}}
\newcommand{\eq}[1]{Eq.~(\ref{eq:#1})}
\newcommand{\secref}[1]{Section~\ref{sec:#1}}
\newcommand{\tab}[1]{Table~\ref{tab:#1}}

\graphicspath{ {images/} }

\begin{document}

\title{Initial data for high-compactness black hole--neutron star binaries}

\author{Katherine Henriksson} \Cornell
\author{Fran\c{c}ois Foucart} \CITA\LBL
\author{Lawrence E. Kidder} \Cornell
\author{Saul A. Teukolsky} \Cornell

\date{\today}

\begin{abstract}
For highly compact neutron stars, constructing numerical initial
data for black hole--neutron star binary evolutions is very
difficult. We describe improvements to an earlier method that
enable it to handle these more challenging cases.  We examine
the case of a 6:1 mass ratio system in inspiral close to merger,
where the star is governed by a polytropic $\Gamma=2$, an SLy, or
an LS220 equation of state. In particular, we are able to obtain
a solution with a realistic LS220 equation of state for a
star with compactness \num{0.26} and mass \SI{1.98}{\Msun}, which
is representative of the highest reliably determined neutron
star masses.  For the SLy equation of state, we can obtain solutions
with a comparable compactness of 0.25, while for a family of
polytropic equations of state, we obtain solutions with compactness
up to \num{0.21}, the largest compactness that is stable in this
family. These compactness values are significantly higher than
any previously published results.
We find that improvements in adapting the computational
domain to the neutron
star surface and in accounting for the center of mass drift of the
system are the key ingredients allowing us to obtain these solutions.
\end{abstract}

\maketitle

\section{Introduction}

Among the prime candidate sources for ground-based gravitational wave
detectors are binary systems containing
inspiraling neutron stars: black hole--neutron
star (BHNS) binaries, or systems containing two neutron stars. Besides
being likely gravitational wave sources, such systems are the leading
candidates to explain short gamma-ray bursts
\cite{eichler:89,1992ApJ...395L..83N,moch:93,LK:98,Janka1999}.
Radioactive decay of the neutron-rich material ejected by the merger
may also power optical/infrared transients days after the
merger~\cite{Roberts2011,metzger:11,2014ApJ...780...31T}, particularly for BHNS binaries
with a large neutron star and a rapidly rotating
black hole~\cite{Foucart:2013a,2013PhRvD..88d1503K,Lovelace:2013vma,Deaton2013,Foucart:2014nda}, and for binary
neutron star mergers with compact neutron stars~\cite{hotokezaka:13}.

Gravitational waves from coalescing binaries are searched for and
analyzed using the matched-filtering
technique~\cite{Finn:1992,Finn1993,Abadie:2011kd,Babak:2012zx}, which
compares the detector output with a bank of templates that model the
waves emitted by the source.  Therefore accurate knowledge of the
expected waveforms of incoming signals is required.
While post-Newtonian templates are expected to be accurate
when the binary is widely separated, they break down near merger.
Fully relativistic
numerical simulations of the last few orbits and the merger
are needed to match onto the post-Newtonian waveforms. Moreover,
modeling of the subsequent electromagnetic and neutrino emission
must also be done by a code that can deal with
all the effects of strong-field gravity.

Numerical modeling of these systems is very challenging (see
\cite{Duez:2009yz,shibata:2011,lrr-2012-8,Lehner2014} for reviews).
A key ingredient in such simulations is accurate initial data.
Ideally, one would like a snapshot of the gravitational field
and the matter distribution only a few orbits before merger
but resulting from millions of years of slow inspiral. In general
relativity, no exact way is known to do this because the nonlinear
Einstein equations are too difficult to solve.
So instead,
various plausible approximations are made.
The most common assumption
is that the binary has had time to settle into a
quasi-equilibrium state, the system being approximately
time-independent in the corotating frame. Furthermore,
as the viscous forces within the star are expected to be
small, we do not expect much change in the spin of the
star as the orbital radius decreases. For an initially nonspinning
neutron star, this would lead to an irrotational
velocity profile, another standard assumption. Because
of gravitational wave emission, however, there is no exact equilibrium
state. Accordingly, these conditions cannot all be perfectly satisfied
simultaneously.
Nevertheless, initial data incorporating these assumptions seems
to work quite well in practice.

This paper will focus on initial data for BHNS binaries, and in particular
systems where the neutron star has high compactness
\begin{equation}
\mathcal{C}=\frac{M_\mathrm{NS}}{R}.
\end{equation}
Here $M_\mathrm{NS}$ is the ADM mass
and $R$ is the areal radius for an isolated star
with the same equation of state and baryon mass, and we use units with
$c=G=1$.
The techniques introduced here should be equally applicable to
neutron star--neutron star binaries.

Previous results on initial data for BHNS evolutions include the early
work of Taniguchi et al.~\cite{Taniguchi:2005fr} and
Sopuerta et al.~\cite{Sopuerta2006}, as well as more recent initial
configurations generated by Taniguchi
et al.~\cite{TaniguchiEtAl:2006,Taniguchi2007,Taniguchi2008} and
Grandclement~\cite{Grandclement2006}. Both
Taniguchi and Grandclement use codes based on the LORENE
package~\cite{LORENE}, and excise from the computational domain
the region inside the apparent horizon of the black hole. An alternative method
based on the puncture formalism, in which the constraints are solved both inside
and outside the black hole horizon, has been proposed by Kyutoku et al.~\cite{Kyutoku:2009}.
The results from~\cite{Kyutoku:2009} were also obtained using LORENE.
A newer initial data code, COCAL, has
been developed by Ury\=u and Tsokaros~\cite{Uryu2012}, but not
yet applied to BHNS binaries.

Our own group has developed an independent code (\cite{FoucartEtAl:2008},
henceforth Paper I) that uses a multidomain spectral method to
achieve high accuracy at a relatively low computational cost.
The code is based on the spectral
elliptic solver (Spells) developed by the Cornell-Caltech
collaboration ~\cite{Pfeiffer2003b}, and originally developed by
Pfeiffer~\cite{Pfeiffer2002a,Pfeiffer2003} for the study of binary
black hole initial data. While the mathematical formulation
of the problem is very similar to~\cite{Taniguchi2008}
and~\cite{Grandclement2006}, the numerical techniques are quite different.
In particular, while all use multidomain spectral methods,
the flexibility that Spells offers in choosing subdomain shapes and the form
of the elliptic equations allows one to
efficiently adapt the configuration of the numerical grid to the geometry
of the system and obtain high-precision
results at a very reasonable computational cost.

A drawback of the method described in Paper I is that it fails to converge
to a solution when the compactness $\mathcal{C}$ of the neutron star
is too high. In fact, this seems to be a defect of all the published
methods for solving the BHNS initial value problem.
The maximum compactness that can be handled depends on the
equation of state (EOS). The easiest EOS for all methods is
a $\Gamma=2$ polytrope because it is smooth
inside the star, and the density goes linearly to zero near the surface.
The method of Paper I can reliably produce binaries with an initial
separation of $7M_0$ and a compactness up to 0.18. Here $M_0$ is defined via
\begin{equation}
    \label{eq:M0def}
    M_0=M_\mathrm{BH}+M_\mathrm{NS}
\end{equation}
where $M_{\rm BH}$ is the Christodoulou mass of the black hole
and $M_{\rm NS}$ is the neutron star mass as defined above.
With some small modifications,
described later, it can reach $\mathcal{C}=0.20$, very close to the
maximum allowed value before the neutron star is unstable to gravitational
collapse.

Treating realistic equations of state is more difficult. They tend
to have nonsmooth behavior as the composition changes in various
density regimes. They often have nonanalytic behavior or very steep
slopes at the surface. And even smooth equations of state may be
given in tabular form, which introduces its own nonsmoothness.
For example, for the SLy EOS~\cite{Douchin:2001,Haensel:2004,Shibata:2005ss}, the maximum compactness
attainable by the method of Paper I is 0.16, corresponding to a neutron
star mass of \SI{1.27}{\Msun}.
The other codes for producing BHNS initial data have reported a maximum
compactness of $\mathcal{C}=0.196$
with a piecewise-polytropic equation of state (for a $1.45M_\odot$ neutron star)
\cite{Kyutoku:2011vz}, while binary neutron star initial data has been
obtained up
to $\mathcal{C}=0.213$
(for a $1.6M_\odot$ neutron star)~\cite{hotokezaka:13}.
Since a neutron star of mass \SI{2}{\Msun}
is known to exist, this is clearly not adequate.

In this paper, we describe several technical improvements to the
algorithm of Paper I that allow high-compactness initial data to
be calculated. For example, we show that for the SLy EOS,
a solution with
$\mathcal{C}=0.25$, corresponding a neutron star mass of \SI{1.86}{\Msun},
can be obtained. Furthermore, for the LS220 EOS~\cite{Lattimer:1991nc},
we can obtain a solution
with $\mathcal{C}=0.26$, corresponding to a mass of \SI{1.98}{\Msun}.
These initial data can now be used in
binary evolutions to study the effect of high compactness on
the outcome of the merger.

In \cite{CorderoCarrion2009}, it was observed that highly compact
configurations may result in mathematical nonuniqueness of the solution,
and a resolution of this problem was presented in the context of
conformally flat formulations for the evolution equations.
In contrast, the work reported here is not related to nonuniqueness. Rather,
we are dealing with an iterative algorithm that displays poor
convergence as the compactness increases. The improvements to the
algorithm restore good convergence.

In this paper, we first describe in \secref{methods} the methods we use to
improve the solution procedure of Paper I, in particular addressing requirements
on the proper determination of the neutron star surface location and the control
of linear momentum in the system. In \secref{results} we  present
high compactness results using polytropic, SLy, and LS220 EOS.
Finally, in \secref{conclusions} we offer closing remarks on these findings.

\section{Methods}
\label{sec:methods}

Our work is based on an implementation in the Spectral Einstein Code
(\SpEC) of the procedure described
in Paper I. The core of the solution procedure is to solve a set
of elliptic equations to
determine the metric and the velocity potential for the matter
(an ``elliptic solve'').
These elliptic equations contain a number of
auxiliary
variables that determine the physical properties of the
initial data. Some of these are imposed based on analytical
considerations, such as the background metric or boundary conditions, while
others are solved for in between iterations in order to enforce
desired physical properties (mass and spin of the compact objects,
orbital parameters of the binary, total linear momentum of the system),
or computationally convenient properties (alignment of the surface of the
neutron star with the boundary between two subdomains).
This procedure is described in detail in Sec. III.C of Paper I.

In solving the elliptic equations, a
relaxation scheme is used to keep the solution in a convergent regime.
Instead of simply using the new solution at every step, it is combined with the
previous one via
\begin{equation}
    \label{eq:relaxscheme}
    u_\mathrm{new}=\lambda u^*+(1-\lambda) u_\mathrm{old}
\end{equation}
where $\lambda$ is a parameter (the relaxation parameter) that we can choose,
$u$ is a quantity such as the value of the metric at some point,
and $u^*$ is the value of $u$ found by solving the elliptic equations.
This scheme decreases the change in the
quantities at each step, which helps to maintain the convergence of the solver.
In Paper I, we used $\lambda=0.3$.

In this paper, we show that in order to obtain initial data for
high-compactness neutron stars, modifications to the iterative procedure
of Paper I
are required.
In particular, we need to modify how we enforce that the location of the surface
of the neutron star is at a subdomain boundary
and that the total linear ADM momentum of the system is zero.

\subsection{Neutron Star Surface Adjustment}
\label{sec:nssurface}

\begin{figure}
    \includegraphics[width=\linewidth]{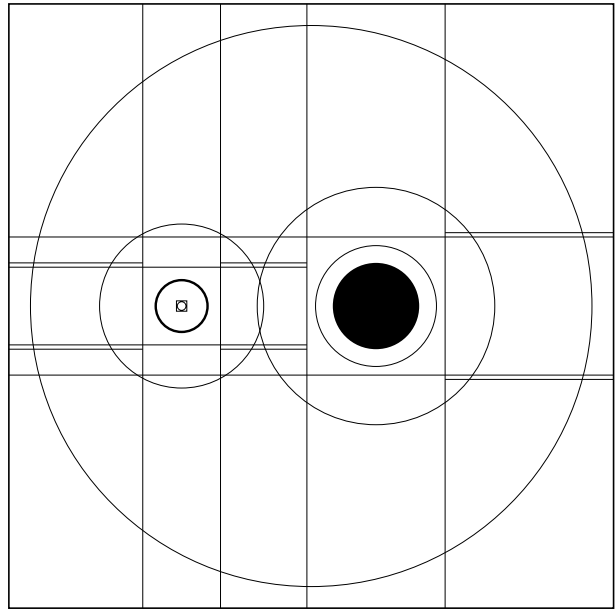}
    \caption{The domain used for the binary black hole--neutron star system.
    The horizon of the black hole is an external boundary of the domain; its
    interior is excised. The thick line shows the star surface; this and all
    other interior lines form subdomain boundaries.
    The largest circle is the inner
    boundary of the outer subdomain, this and all circles show the boundaries
    of spherical shells. The domain furthermore comprises eight
    cylindrical shells, in two stacked groups of five and three, and three
    rectangular prisms between and beside the objects.}
    \label{fig:thedomain}
\end{figure}

An important aspect of the solver is the numerical domain that is used to solve
the equations.
We employ the domain described in Paper I Section III.A, and
shown here in \fig{thedomain}.
The physical domain is covered by a number of
overlapping spectral subdomains;
choosing the location and resolution of the subdomains
allows the domain to be well-matched to the problem. A particularly important
detail (we have found)
is that since the neutron star surface is a physical discontinuity,
it should be placed at a subdomain boundary
to avoid Gibbs oscillations in the numerical solution.
This requirement was also found in Paper I,
although we used a different technique to enforce it.
We employ a modification of the method described in Paper I in Section III.A.2
and in item 3 of the iterative procedure of Section III.C.

We define the neutron star surface as the location where the enthalpy reaches
some target value. The
enthalpy $h$ is defined via $h=1+\epsilon+P/\rho$, with $\rho$ the
baryon density, $\epsilon$ the internal energy density, and $P$ the pressure.
In practice, the enthalpy is computed from the metric and the fluid velocity
by assuming hydrostatic equilibrium (see Paper I).
Since the neutron star surface may deviate slightly from the subdomain
boundary during the solution,
the equation for $h$ is solved
outside the star as well as inside. In principle, the enthalpy
should take on a constant value
everywhere in the vacuum region, but
numerically $h$ is allowed to
take values below $h(\rho=0)$, which ensures better
behavior in the surface-finding algorithm. Similarly, the baryon density
$\rho$ is allowed to take unphysical values $\rho<0$ in regions in which
$h<h(\rho=0)$ to guarantee that $\nabla \rho$ is continuous---a desirable
property when using spectral methods.

The location of the surface can be found with a simple one-dimensional root
finding algorithm along the collocation directions,
which lets us find the stellar radius $R(\theta,\phi)$ as a
function of angle. This is then used to define the stellar surface coefficients
via an expansion in scalar spherical harmonics:
\begin{equation}
    R(\theta,\phi)=\sum_{l}\sum_{m=-l}^l R_{lm} Y_l^m(\theta,\phi)~.
\end{equation}
Note that for $m\ne 0$, the coefficients $R_{lm}$ will be complex-valued in
general, but because $R(\theta,\phi)$ is real-valued, $R_{l,-m} = (-1)^m (R_{lm})^\star$
and so we can define
\begin{equation}
    \label{eq:Slmdef}
    S_{lm}=\begin{cases}
    (-1)^m \sqrt{\frac{2}{\pi}} \Re(R_{lm}) & m \ge 0 \\
    (-1)^m \sqrt{\frac{2}{\pi}} \Im(R_{lm}) & m < 0
    \end{cases}
\end{equation}
to store all independent components of the $R_{lm}$ coefficients.

The coefficients in \eq{Slmdef}
are used to update the domain boundary for the neutron star subdomains
used by the solver.
However, since the enthalpy may have non-negligible errors early in
the solve, the computed surface may not be in the right place and may
have large jumps between iterations. This is similar to the difficulties that
are encountered in the elliptic solve.
Accordingly,
we follow a relaxation scheme as in \eq{relaxscheme} in updating the
$S_{lm}$ coefficients defined in \eq{Slmdef}, which are
computed in each step and used to define the mapping
shown in Eqs.\ (77) and (78) in Paper I. We have found
that choosing the same value
for this relaxation parameter as for the one controlling the metric and matter
relaxation gives good results.

\subsection{ADM Momentum Control}

To uniquely specify the initial conditions that we are attempting to produce,
we need to fix the location of the centers of the black hole and neutron star,
${\bf c}_{\rm BH}$ and ${\bf c}_{\rm NS}$, the initial orbital angular velocity
of the binary $\Omega_0$, and its radial velocity $v_r=\dot{a_0}r$.
We also have the freedom to choose
the value of the shift vector on the outer boundary of the computational domain,
${\bf \beta}(r_{\rm out})={\bf v}^{\rm boost}$. We try to make these choices so
that
\begin{itemize}
\item The linear ADM momentum of the system satisfies ${\bf P}^{\rm ADM}=0$.
\item The objects are following
circular orbits with the desired initial separation $d$.
\item The center of mass of the system is at the origin of the chosen coordinate system.
\end{itemize}

The initial data solver finds constraint-satisfying configurations by following the iterative
procedure described in Section III.C of Paper I.
The center of the neutron star is
fixed at ${\bf c}_{\rm NS}=(-d\, M_{\rm BH}/M_0,0,0)$.
The quantities $\Omega_0$ and $\dot{a_0}$ are
chosen in order to minimize the orbital eccentricity of the system, following the iterative
procedure developed for black hole--black hole binaries~\cite{Pfeiffer-Brown-etal:2007}.
Alternatively, we can solve for $\Omega_0$ by requiring force balance at the center of
the neutron star, following Eq.\ (48) of Paper I as in step 5 of the
iterative method described in
Section III.C of Paper I. Combined with the choice $\dot{a_0}=0$,
this leads to eccentricities of a few percent, and provides a good initial guess for the
eccentricity reduction algorithm. This leaves us with the choices of ${\bf c}_{\rm BH}$
and ${\bf v}^{\rm boost}$, which are both made iteratively. The location of the black hole
center can be modified at each step of the iteration, after we solve the constraint equations
and evaluate the position of the neutron star surface (step 4 of the iterative procedure in
Paper I). The choice of ${\bf v}^{\rm boost}$ comes as an outer boundary condition in the constraint
equation for the shift. We have developed two different methods
to choose ${\bf c}_{\rm BH}$ and ${\bf v}^{\rm boost}$.

The first (hereafter ``position control'') is largely similar to the algorithm
described in Paper I for spin-aligned binaries, and updated in
\citet{Foucart:2010eq} for black hole spins misaligned
with the orbital angular momentum of the binary. In this method, ${\bf c}_{\rm BH}$ is initialized
to ${\bf c}^0_{\rm BH}=(d\, M_{\rm NS}/M,0,0)$.
At each step $n$ of the iterative procedure, we measure
the linear ADM momentum ${\bf P}_n$, and the relative changes in each component
$\delta P^i = |P^i_n-P^i_{n-1}|/|P^i_n|$. If $\delta P^i<\alpha_P$
for the largest component $P_n^i$
of the ADM momentum and a freely specifiable parameter
$\alpha_P$, then we reset the components of ${\bf c}_{\rm BH}$ in the orbital plane of the binary
(the $x$-$y$ plane here) using the relaxation formula
\begin{equation}
c_{x,y}^{\rm new} = \lambda_P c_{x,y}^* + (1-\lambda_P) c_{x,y}^{\rm old}
\end{equation}
analogously to \eq{relaxscheme}, with $c_{x,y}^*$ computed using the
values of $P^{x,y}$ and $c_{x,y}$ at the two latest
steps $i$ and $j$ at which the location of the black hole center was modified:
\begin{equation}
c_{x,y}^* = \frac{c_{x,y}^i P^{y,x}_j-c_{x,y}^j P^{y,x}_i}{P^{y,x}_j-P^{y,x}_i}.
\label{eq:cstar}
\end{equation}
For larger neutron stars, we have been using $\alpha_P=0.1$, $\lambda_P=1$.
For compact stars, we find that changing
the location of the black hole center more often, but by smaller increments, works better.
Accordingly, we use $\alpha_P=0.4$,
and $\lambda_P$ is chosen to
equal the relaxation parameter in the elliptic solve.
Eq.~(\ref{eq:cstar}) is inspired by the fact that, in Newtonian physics, a change $\delta{\bf c}$
in ${\bf c}_{\rm BH}$ induces a change $\delta {\bf P}=-\delta{\bf c} \times {\bf \Omega}$ in ${\bf P}$.
A similar updating algorithm is used for the vertical location of the black hole center, except that
instead of trying to cancel the linear momentum of the system, we require vertical force balance $(\nabla h)_z=0$
at the center of the neutron star. Thus we use
\begin{equation}
c_z^* = \frac{c_z^i (\nabla h)_z^j-c_z^j (\nabla h)_z^i}{(\nabla h)_z^j-(\nabla h)_z^i}.
\end{equation}
The vertical component of the linear momentum, which cannot be easily controlled by moving the location
of the compact objects, is instead canceled by a ``boost'' given to the entire system through
the chosen value ${\bf v}^{\rm boost}$ of the shift on the outer boundary. We set $v^{\rm boost}_{x,y}=0$
and update $v^{\rm boost}_z$ using the same method as for the black hole center, but with
\begin{equation}
v^*_z=\frac{v^i_z P^z_j-v^j_zP^z_i}{P^z_j-P^z_i}.
\end{equation}

In the second method (hereafter ``boost control''), the location of the center of the black hole
is fixed to its expected value for a Newtonian binary orbiting around the origin of our coordinate
system, ${\bf c}_{\rm BH}=(d\, M_{\rm NS}/M,0,0)$. The constraint ${\bf P}^{\rm ADM}=0$ is then satisfied
by controlling all components of the linear momentum through the outer boundary condition on the shift.
That is, we use
 \begin{equation}
v^*_{x,y,z}=\frac{v^i_{x,y,z} P^{x,y,z}_j-v^j_{x,y,z}P^{x,y,z}_i}{P^{x,y,z}_j-P^{x,y,z}_i}
\end{equation}
to reset the components of ${\bf v}^{\rm boost}$ whenever $\delta P^i<\alpha_P$.

For spin-aligned binaries, position control generally results in a very small coordinate velocity
for the center of mass of the system. On the other hand, imposing a non-zero boundary condition on
the shift at large distances leads to a drift of the coordinate location of the center of mass at
velocity $v^{\rm COM}\sim v^{\rm boost}$. This effect was observed for misaligned BHNS binaries
in~\cite{Foucart:2010eq}. Because a large drift of the center of mass might introduce undesirable
coordinate effects in the methods used to extrapolate the gravitational wave signal to infinity,
or complicate the work of the control system used to evolve the binary in the comoving frame, we
have generally preferred position control. However, each change of the location of the center of
the black hole in the initial data solver introduces significant constraint violations in our solution.
We find that, for very compact stars with $\mathcal{C}\gtrsim 0.2$,
these changes can prevent convergence of the iterative procedure
used to generate initial data, and so we avoid position control in these
cases.

\section{Results}
\label{sec:results}

Using the methods described above, we were able to obtain solutions
for binaries with various neutron star compactness for a variety of different
equations of state. In all cases, we have chosen the black hole to be
nonspinning, since the aim of the study was to develop methods
for handling high neutron star compactness.

\subsection{Polytropic $\mathbf{\Gamma\boldsymbol=2}$ EOS}

Using the methods of Paper I, tweaking only the
elliptic solve relaxation
parameter described in the iterative procedure in Section III.C, solutions
for polytropic $\Gamma=2$ equations of state
with compactness up to 0.18 could be obtained. If we additionally incorporate
an initial guess based on a lower-compactness binary instead
of starting with an isolated neutron star configuration, a solution with
$\mathcal{C}=0.20$ could be found. Finally,
using the techniques described in \secref{methods}, we are able to obtain
solutions with compactness up to
$\mathcal{C}=0.21$ and mass of \SI{1.4}{\Msun}.
For the family of equations of
state we consider, there is a dynamic instability that sets in for
compactness slightly higher than 0.21, and so this is roughly the highest
compactness that we expect to be physically meaningful. In addition, other work
\cite{CorderoCarrion2009}
has found mathematical nonuniqueness problems with obtaining solutions for
stars that are
dynamically unstable, and for both these reasons we avoid investigating such
solutions.
For polytropes, we could thus nearly reach the maximum stable compactness
without further modifications of the elliptic solve.
Only the highest stable
compactness $\mathcal{C}=0.21$ still required the techniques introduced above.
By contrast,
for the SLy and LS220 equations of state,
the methods of Paper I fail at much lower compactness.

\begin{figure*}
    \includegraphics{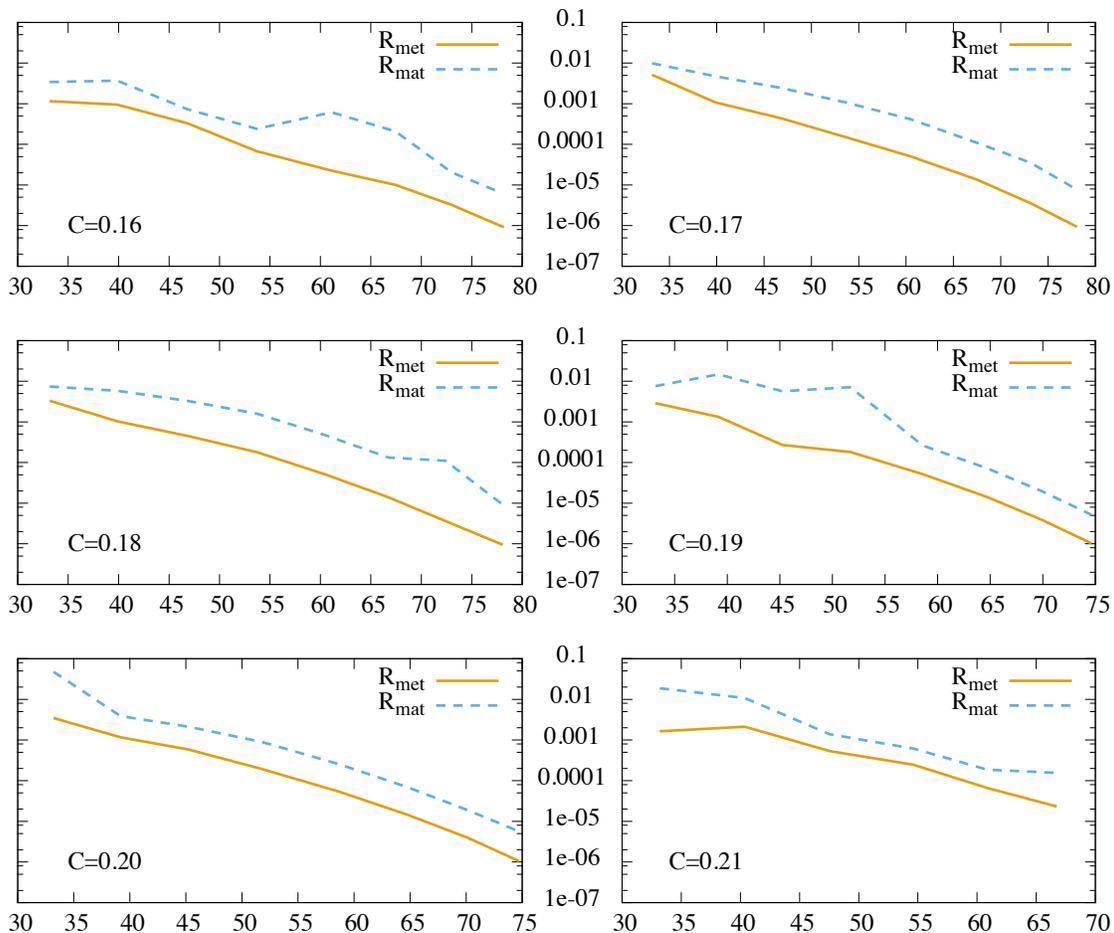}
    \caption{Spectral convergence for a family of BHNS binary configurations.
    For each configuration, the equation of state is a $\Gamma=2$ polytrope
    chosen to yield $M_\mathrm{NS}=\SI{1.4}{\Msun}$.
    Each panel shows the final residual in the
    metric solution ($R_\mathrm{met}$) and in the
   velocity potential solution
    ($R_\mathrm{mat}$) as a function of the cube root of the number of grid
    points in the domain. The vertical scale is logarithmic, and the exponential
    convergence with resolution is apparent.
    }
    \label{fig:polyconv}
\end{figure*}

We show in \tab{polysolve} the results for applying this method to $\Gamma=2$
polytropes of different compactness. The stellar mass is held constant at
\SI{1.4}{\Msun} and so the polytropic parameter $\kappa$ varies across these
solutions and is included in the table. (Note that for polytropes,
the mass can be rescaled by changing $\kappa$ without affecting
$\mathcal{C}$.)
The mass ratio is 6:1 for all cases.
The binding energy $E_b$ is computed
by subtracting the sum of the ADM masses of the black hole and
neutron star in isolation from the ADM energy of the system.
The surface coefficients shown are defined in \eq{Slmdef}.
The residuals obtained as a function of
resolution are shown in \fig{polyconv}. In the residuals we can see some
aberrant behavior and jumps, but nevertheless the exponential
convergence with resolution that characterizes spectral methods is apparent,
especially if one focuses on the higher resolutions. The convergence with
resolution also does not seem to vary among the different cases,
so that compactness does not seem to be an issue here. One other thing to note
here is that the initial distance $d$ between the black hole and neutron
star is chosen to be quite close in all but the highest compactness cases,
as compared with the values in Tables 2, 5, and 6 in Paper I.

\begin{table}

    \begin{tabular}{|c|c|c|c|c|c|c|c|}
        \hline
        $\mathcal{C}$ & $\kappa$ & $\lambda$ & $d/M_0$
        & $\Omega M_0$ & $E_b/M_0$
        & $J/{M_0}^2$ & $S_{20}/S_{00}$ \\ \hline
        0.15 & 96.7 & 0.3 & 6.86 & 0.04648 & -6.69 & 0.422 & 4.261 \\
        0.16 & 89.7 & 0.3 & 6.86 & 0.04649 & -6.70 & 0.421 & 4.439 \\
        0.17 & 84.1 & 0.2 & 6.86 & 0.04650 & -6.71 & 0.421 & 4.474 \\
        0.18 & 79.8 & 0.2 & 6.86 & 0.04650 & -6.71 & 0.421 & 4.412 \\
        0.19 & 76.6 & 0.2 & 7.71 & 0.03967 & -6.19 & 0.434 & 3.574 \\
        0.20 & 74.4 & 0.2 & 7.71 & 0.03968 & -6.19 & 0.434 & 3.419 \\
        0.21 & 73.2 & 0.15 & 14.0 & 0.01740 & -3.85 & 0.526 & 1.441 \\
        \hline \multicolumn{8}{l}{} \\[-2ex]
        \multicolumn{5}{l}{} & \multicolumn{1}{c}{$\times10^{-3}$} &
        \multicolumn{1}{l}{} & \multicolumn{1}{c}{$\times10^{-3}$}
    \end{tabular}
    \caption{Solved quantities for a family of BHNS binary configurations. For
    each configuration, the equation of state is a $\Gamma=2$ polytrope chosen
    to yield $M_\mathrm{NS}=\SI{1.4}{\Msun}$.
    The multiplier below applies to the column above it.
    The columns show the compactness
    $\mathcal{C}$, the polytropic parameter $\kappa$,
    the relaxation parameter $\lambda$, the initial separation $d$,
    the orbital angular velocity $\Omega$, the
    binding energy $E_b$, the ADM angular momentum $J$, and the ratio
    $S_{20}/S_{00}$ of surface coefficients defined in \eq{Slmdef}.
    The quantity $M_0$ is defined in \eq{M0def}.
   In all cases the black hole has no spin and the mass ratio is 6:1.
    }
    \label{tab:polysolve}

\end{table}

\subsection{SLy $\mathbf{\Gamma\boldsymbol=2}$ EOS}

Using the methods of this paper, we can obtain solutions
for an SLy equation of state with compactness up to
$\mathcal{C}=0.25$, which corresponds to a star with a mass of \SI{1.86}{\Msun}.
For comparison, the maximum stable compactness for this equation of
state is about $\mathcal{C}=0.31$, corresponding to a mass of \SI{2.04}{\Msun}.
Solutions were found for a variety of different distances,
although as the compactness
becomes greater, it becomes more difficult to obtain solutions for
very close binaries.
At the highest compactness, the closest binaries we attempted
did not have a convergent solution, and so an evolution of such a system would
need to simulate more than 20 orbits before a merger.

\begin{figure*}
    \includegraphics{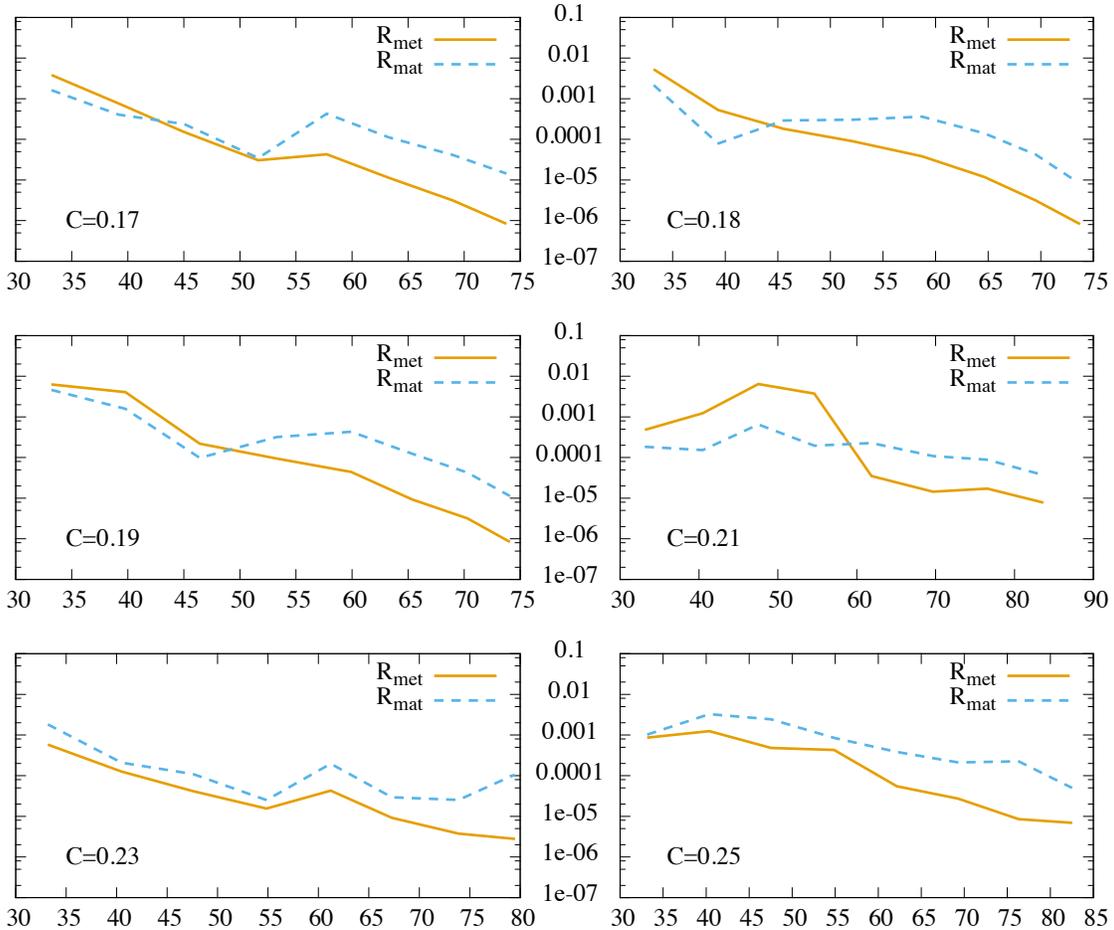}
    \caption{Spectral convergence for a family of BHNS binary configurations.
    For each configuration, the equation of state is SLy.
    Each panel shows the final residual in the
    metric solution ($R_\mathrm{met}$) and in the
  velocity potential solution
    ($R_\mathrm{mat}$) as a function of the cube root of the number of grid
    points in the domain. The vertical scale is logarithmic, and the exponential
    convergence with resolution is apparent.
    }
    \label{fig:slyconv}
\end{figure*}

These results are shown in \tab{slysolve}. The mass ratio $q$
is close to 6, with slight variations
between solutions.
The residuals obtained in
the solve are shown in \fig{slyconv}, plotted against $N^{1/3}$, the cube root
of the number of points in the domain. Note that for the $\mathcal{C}=0.23$
and $\mathcal{C}=0.25$ cases, we have only plotted the convergence in the case
of $d=14M_0$. Even more so than in the polytropic case,
there are some jumps in the residuals, particularly in the
velocity potential. Furthermore, jumps are present at all
compactnesses instead of just low ones. However, once again, the convergence
does appear smoothly exponential at high resolution. The same value for the
relaxation parameter is used for all of the various quantities that are updated
via a relaxation scheme.

\begin{table}

    \begin{tabular}{|c|c|c|c|c|c|c|c|c|}
        \hline
        $\mathcal{C}$ & $M_\mathrm{NS}$ & $q$ & $\lambda$ & $d/M_0$
        & $\Omega M_0$ & $E_b/M_0$ & $J/{M_0}^2$
        & $S_{20}/S_{00}$ \\ \hline
        0.16 & 1.27 & 6.64 & 0.2 & 6.95 & 0.04560 & -6.15 & 0.393 & 4.928 \\
        0.17 & 1.34 & 6.26 & 0.2 & 6.90 & 0.04611 & -6.46 & 0.409 & 4.696 \\
        0.18 & 1.41 & 5.94 & 0.2 & 6.85 & 0.04659 & -6.77 & 0.424 & 4.460 \\
        0.19 & 1.49 & 5.65 & 0.2 & 6.80 & 0.04707 & -7.05 & 0.438 & 4.231 \\
        0.21 & 1.62 & 6.0 & 0.15 & 13.71 & 0.01791 & -3.91 & 0.522 & 1.533 \\
        0.23 & 1.75 & 6.0 & 0.15 & 8.0 & 0.03785 & -6.05 & 0.438 & 2.878 \\
        0.23 & 1.75 & 6.0 & 0.15 & 10.0 & 0.02785 & -5.09 & 0.468 & 2.066 \\
        0.23 & 1.75 & 6.0 & 0.15 & 12.0 & 0.02160 & -4.39 & 0.497 & 1.611 \\
        0.23 & 1.75 & 6.0 & 0.15 & 14.0 & 0.01739 & -3.82 & 0.526 & 1.324 \\
        0.23 & 1.75 & 6.0 & 0.15 & 16.0 & 0.01439 & -3.22 & 0.552 & 1.124 \\
        0.25 & 1.86 & 6.0 & 0.1 & 14.0 & 0.01738 & -3.88 & 0.525 & 1.144 \\
        0.25 & 1.86 & 6.0 & 0.1 & 16.0 & 0.01438 & -2.38 & 0.552 & 0.971 \\
        \hline \multicolumn{8}{l}{} \\[-2ex]
        \multicolumn{6}{l}{} & \multicolumn{1}{c}{$\times10^{-3}$} &
        \multicolumn{1}{l}{} & \multicolumn{1}{c}{$\times10^{-3}$}
    \end{tabular}
    \caption{Solved quantities for a family of BHNS binary configurations.
    For each configuration, the equation of state is SLy.
    The multiplier below applies to the column above it.
    The quantities shown are all as for \tab{polysolve}, with $\kappa$
    replaced by the neutron star ADM mass $M_\mathrm{NS}$ in units
    of $\Msun$.
    We also show the mass ratio $q$ for each configuration.}
    \label{tab:slysolve}

\end{table}

\subsection{LS220 EOS}

We obtain solutions for an LS220 tabulated equation of state with compactness
up to $\mathcal{C}=0.26$, which corresponds to a star with a mass of
\SI{1.98}{\Msun}. For comparison, the largest known reliable neutron
star masses are
\SI{1.97\pm 0.04}{\Msun} \cite{Demorest:2010bx} and \SI{2.01\pm 0.04}{\Msun}~\cite{2013Sci...340..448A}.
The LS220 equation of state is the most realistic of
the ones considered here, and so this is a very relevant result.
The largest compactness for which this equation of state
yields a stable solution is $\mathcal{C}=0.29$, corresponding to a mass of
\SI{2.04}{\Msun}.

\begin{figure*}
    \includegraphics{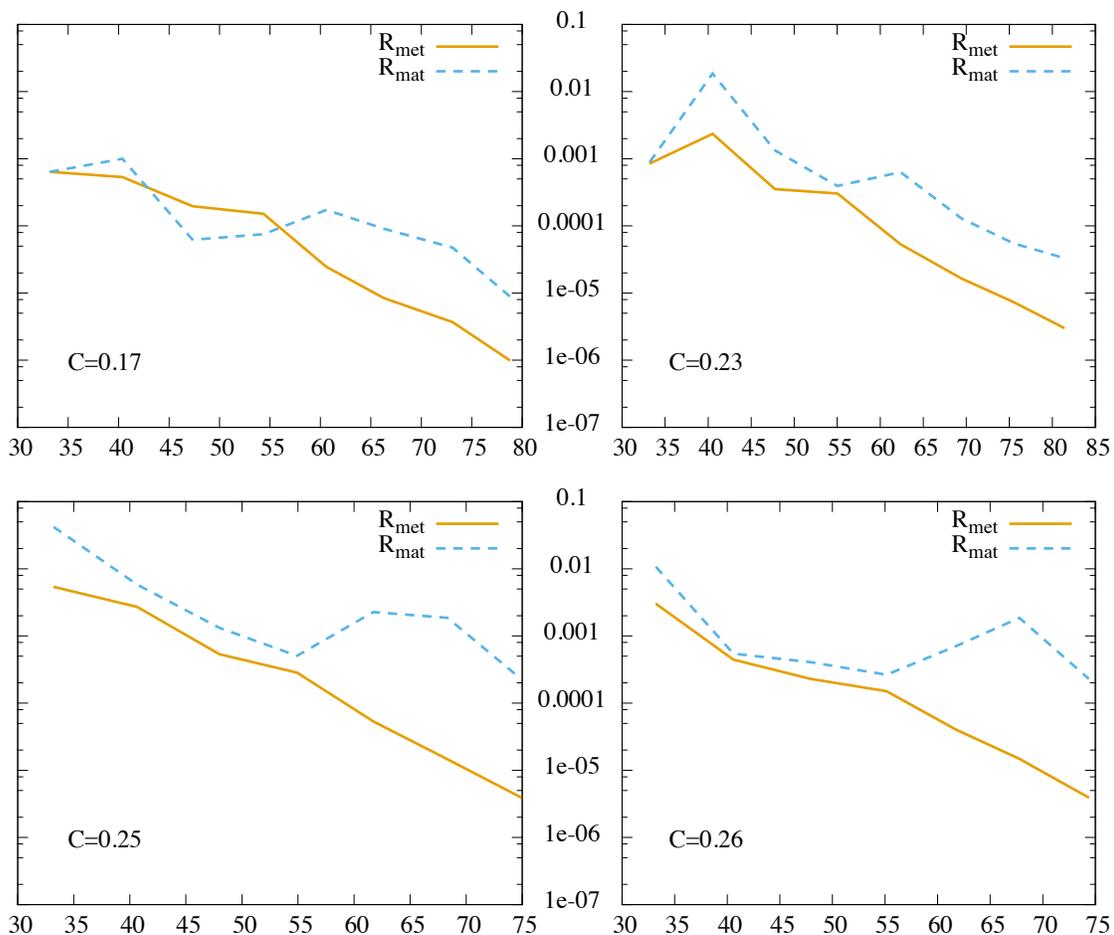}
    \caption{Spectral convergence for a family of BHNS binary configurations.
    For each configuration, the equation of state is LS220.
    Each panel shows the final residual in the
    metric solution ($R_\mathrm{met}$) and in the
    velocity potential solution
    ($R_\mathrm{mat}$) as a function of the cube root of the number of grid
    points in the domain. The vertical scale is logarithmic, and the exponential
    convergence with resolution is apparent.
    }
    \label{fig:ls220conv}
\end{figure*}

These results are shown in \tab{ls220solve}.
In all cases, the mass ratio is 6:1.
The residuals obtained in
the solve are shown in \fig{ls220conv}, plotted against $N^{1/3}$, the cube root
of the number of points in the domain. As before, we find exponential
convergence, and obtain a final residual of \num{e-6} or nearly so.

\begin{table}

    \begin{tabular}{|c|c|c|c|c|c|c|c|}
        \hline
        $\mathcal{C}$ & $M_\mathrm{NS}$ & $\lambda$ & $d/M_0$
        & $\Omega M_0$ & $E_b/M_0$ & $J/{M_0}^2$
        & $S_{20}/S_{00}$ \\ \hline
        0.17 & 1.40 & 0.2 & 14.0 & 0.01739 & -3.84 & 0.526 & 1.885 \\
        0.23 & 1.83 & 0.1 & 14.0 & 0.01739 & -3.83 & 0.526 & 1.347 \\
        0.25 & 1.94 & 0.08 & 16.0 & 0.01439 & -3.18 & 0.552 & 0.985 \\
        0.26 & 1.98 & 0.07 & 14.0 & 0.01739 & -3.73 & 0.525 & 1.075 \\
        \hline \multicolumn{8}{l}{} \\[-2ex]
        \multicolumn{5}{l}{} & \multicolumn{1}{c}{$\times10^{-3}$} &
        \multicolumn{1}{l}{} & \multicolumn{1}{c}{$\times10^{-3}$}
    \end{tabular}
    \caption{Solved quantities for a family of BHNS binary configurations.
    For each configuration, the equation of state is the LS220 tabulated
    equation of state.
    The multiplier below applies to the column above it.
    The quantities shown are all as for \tab{polysolve}, with $\kappa$
    replaced by the neutron star ADM mass $M_\mathrm{NS}$ in units
    of $\Msun$.}
    \label{tab:ls220solve}

\end{table}

\section{Conclusions}
\label{sec:conclusions}

The problem of solving for initial data for a black hole--neutron star system
is a complex one involving many interacting solution steps. In addition
to the elliptic equations governing the system, multiple nonlinear
constraints must be imposed simultaneously. In this paper, we focused
on two of these in particular: aligning the surface with a subdomain
boundary and enforcing zero linear momentum on the binary system.
The close interaction between the equations and all of the constraints requires
care in solving the system, and for some choices of system parameters the
convergence of the system can be very sensitive to deficiencies in the solution
method. In fact, it seems to be generally true that high compactness stars
cause difficulty with the various solvers that exist.

We have found that by applying additional relaxation steps
to the method of Paper I \cite{FoucartEtAl:2008}
we could obtain solutions for a wide variety of systems
with quite high compactness stars, including stars with polytropic, SLy, and
LS220 equations of state. This includes a solution for a binary with a
physically realistic LS220 equation of state star having a mass of
$M=\SI{1.98}{\Msun}$. We have
found it effective to employ a relaxation scheme in updating the metric and
velocity potential quantities from the elliptic solver
procedure, as well as the neutron
star surface location, in
addition to a boost parameter in a control scheme adjusting for
undesired linear ADM momentum in the system. The choice of different relaxation
parameters allows one to tune the solver for the convergence properties of the
particular solution being examined. These improvements allow
solutions to the black hole--neutron star initial data problem to be
found for physically important cases.

\begin{acknowledgments}
K.H.\ would like to thank Geoffrey Lovelace, Curran Muhlberger, Harald
Pfeiffer, and David Chernoff for useful discussions, and Andy Bohn for
the use of computing resources.
This work was supported in part by
NSF Grants PHY-1306125 and AST-1333129 at Cornell University, and
by a grant from the Sherman Fairchild Foundation.
F.F.\ gratefully acknowledges support from the Vincent and Beatrice
Tremaine Postdoctoral Fellowship and NSERC Canada.
Support for this work was provided by NASA through Einstein Postdoctoral
Fellowship grant number PF4-150122 awarded by the Chandra X-ray Center, which is operated by
the Smithsonian Astrophysical Observatory for NASA under contract NAS8-03060.
This research
was performed in part using the Zwicky computer system operated
by the Caltech Center for Advanced Computing Research
and funded by NSF MRI No. PHY-0960291 and the
Sherman Fairchild Foundation.

\end{acknowledgments}

\bibliography{References/References}

\end{document}